\documentclass[twocolumn,prl,showpacs,preprintnumbers,amsmath,amssymb]{revtex4-1}
\usepackage[english]{babel}
\usepackage{ucs}
\usepackage[utf8x]{inputenc}
\usepackage{graphicx}
\usepackage{amssymb}
\begin{document}

\author{J. Kotakoski$^1$, A.~V. Krasheninnikov$^{1,2}$, U. Kaiser$^3$ and J.~C. Meyer$^{3,}$}
\altaffiliation{Current address: University of Vienna, Department of Physics, Strudlhofgasse 4, 1090 Vienna, Austria} 
\affiliation{$^1$ Department of Physics, University of Helsinki,
  P.O. Box 43, 00014 Helsinki, Finland\\$^2$ Department of Applied
  Physics, Aalto University, P.O. Box 1100, 00076 Aalto, Finland\\$^3$
  Electron microscopy of materials science, University of Ulm,
  Germany}

\title{From Point Defects in Graphene to Two-Dimensional
  Amorphous Carbon}

\begin{abstract}
  While crystalline two-dimensional materials have become an
  experimental reality during the past few years, an amorphous 2-D
  material has not been reported before. Here, using electron
  irradiation we create an $sp^{2}$-hybridized one-atom-thick flat
  carbon membrane with a { \it random } arrangement of polygons,
  including four-membered carbon rings. We show how the transformation
  occurs step-by-step by nucleation and growth of low-energy
  multi-vacancy structures constructed of rotated hexagons and other
  polygons. Our observations, along with first-principles
  calculations, provide new insights to the bonding behavior of carbon
  and dynamics of defects in graphene. The created domains possess a
  band gap, which may open new possibilities for engineering
  graphene-based electronic devices.
\end{abstract}


\maketitle

Hexagonal rings serve as the building blocks for the growing number of
$sp^2$--bonded low-dimensional carbon structures such as
graphene~\cite{Novoselov2004,Novoselov2005a} and carbon
nanotubes~\cite{Iijima1991}. Non-hexagonal rings usually lead to the
development of non-zero curvature, {\it e.g.}, in
fullerenes~\cite{Kroto1987} and carbon nanohorns~\cite{Iijima1999},
where the arrangement of other polygons can be geometrically deduced
via the isolated pentagon rule (IPR)~\cite{Kroto1987,Schmalz1988} and
Euler's theorem~\cite{Terrones1992}. Aberration corrected
high-resolution transmission electron microscopy (AC-HRTEM) has
recently allowed atomic-resolution imaging of regular carbon
nanostructures and identification of defects in these
materials~\cite{Hashimoto2004,Meyer2008,Gass2008,Girit2009,Warner2009}.
Point defects, mostly vacancies, are naturally created by the
energetic electrons of a TEM. However, the possibility for selectively
creating topological defects representing agglomerations of
non-hexagonal rings could be more desirable in the context of
carbon-based electronics.~\cite{CastroNeto2009,Lahiri2010,Yazyev2010b}

In fact, despite the recent advances, the precise microscopic picture
of the response of graphene to electron irradiation remains
incomplete. Earlier experiments on curved carbon
nanosystems~\cite{Sun2006,Sun2008} have shown that they avoid
under-coordinated atoms under irradiation at high temperatures via
vacancy migration and coalescence. Recent experiments on
graphene~\cite{Warner2009} reported only the development of
holes. Theoretical studies have also predicted the appearance of small
holes or formation of h{\ae}ckelite-like
configurations~\cite{Terrones2000} or dislocations~\cite{Jeong2008}.

\begin{figure*}[ht!]
\includegraphics[width=1\linewidth]{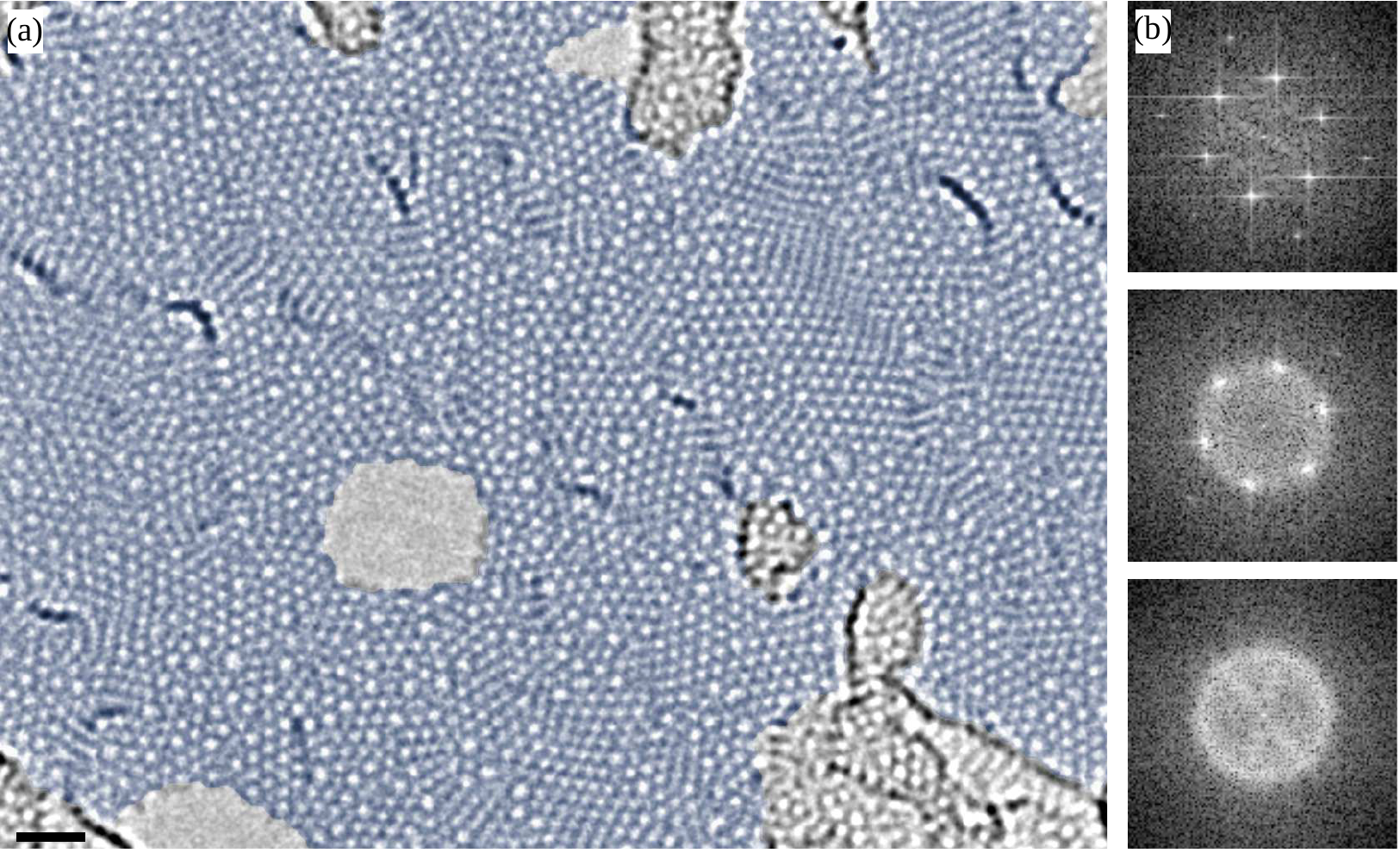}
\caption{(a) (Color online) Amorphous two-dimensional $sp^{2}$-bonded
  carbon membrane created by a high-dose exposure of graphene to
  100~keV electron irradiation in an HRTEM. The blue colored area is a
  single-layer carbon structure. Scale bar is 1~nm. (b) Fourier
  transforms (power spectra) from HRTEM images of the initial graphene
  configuration (top), an intermediate configuration (center), and of
  the amorphous 2D carbon (bottom).}
\label{fgr:amorph} 
\end{figure*}

In this Letter, we report the transformation of graphene into a
two-dimensional random arrangement of polygons due to continuous
exposure to the electron beam with an energy just above the knock-on
threshold. By carefully choosing the electron energy, we selectively
enhance and suppress the underlying mechanisms of defect production.
A combination of experiments and density-functional theory (DFT)
calculations allows us to show that the transformation is driven by
two simple mechanisms: atom ejection and bond rotation. The created
defects tend to have a low formation energy and exhibit an electronic
band gap. We also discover other unexpected configurations, such as
stable carbon tetragons~\cite{Legrand2010} which appear upon linear
arrangement of di-vacancies.

Our graphene membranes were prepared by micro-mechanical cleavage and
transfer to TEM grids~\cite{Meyer2008a}. Aberration-corrected HRTEM
imaging was carried out in an FEI Titan 80--300, equipped with an
objective-side image corrector. The microscope was operated at 80~keV
and 100~keV for HRTEM imaging, and at 300~keV for irradiation.  The
extraction voltage of the field emission source was set to a reduced
value of 2~kV in order to reduce the energy spread. For both 80~keV
and 100~keV imaging, the spherical aberration was set to
20~{\AA{}}$\mu$m and images were obtained at Scherzer defocus
(ca. $-9$~nm). At these conditions, dark contrast can be directly
interpreted in terms of the atomic structure.

The DFT calculations were carried out with the VASP simulation
package~\cite{Kresse1996a, Kresse1996} using projector augmented wave
potentials~\cite{Blochl1994} to describe core electrons, and the
generalized gradient approximation~\cite{Perdew1996} for exchange and
correlation. Kinetic energy cutoff for the plane waves was 500~eV, and
all structures were relaxed until atomic forces were below
0.01~eV/{\AA{}}. The initial structure consisted of 200 C atoms, and
Brillouin zone sampling scheme of Monkhorts-Pack~\cite{Monkhorst1976}
with up to $9\times9\times1$ mesh was used to generate the $k$-points.
Barrier calculations were carried out using the nudged elastic band
method as implemented in VASP~\cite{Henkelman2000}.

We started our experiments by monitoring {\em in situ} the behavior of
graphene under a continuous exposure to electron irradiation using
AC-HRTEM imaging with an electron energy of 100~keV, {\it i.e.}, just
above the threshold for knock-on damage ($T_d$) in $sp^{2}$-bonded
carbon structures~\cite{Banhart1999,Smith2001}. Fig.~\ref{fgr:amorph}a
shows a graphene structure after an electron dose of $\sim
1{\cdot10}^{10}$~${e^{-}}/{\textrm{nm}^{2}}$. Contrary to the
expectations, the structure does not predominantly consist of holes or
collapse into a 3D object. Instead, it has remained as a coherent
single-layer membrane composed of a random patch of polygons.  Holes
have also formed, but only on a small fraction of the area. The
Fourier transform of the image shows that the resulting structure is
completely amorphous (Fig~\ref{fgr:amorph}b).

\begin{figure*}[ht!]
\includegraphics[width=1\linewidth]{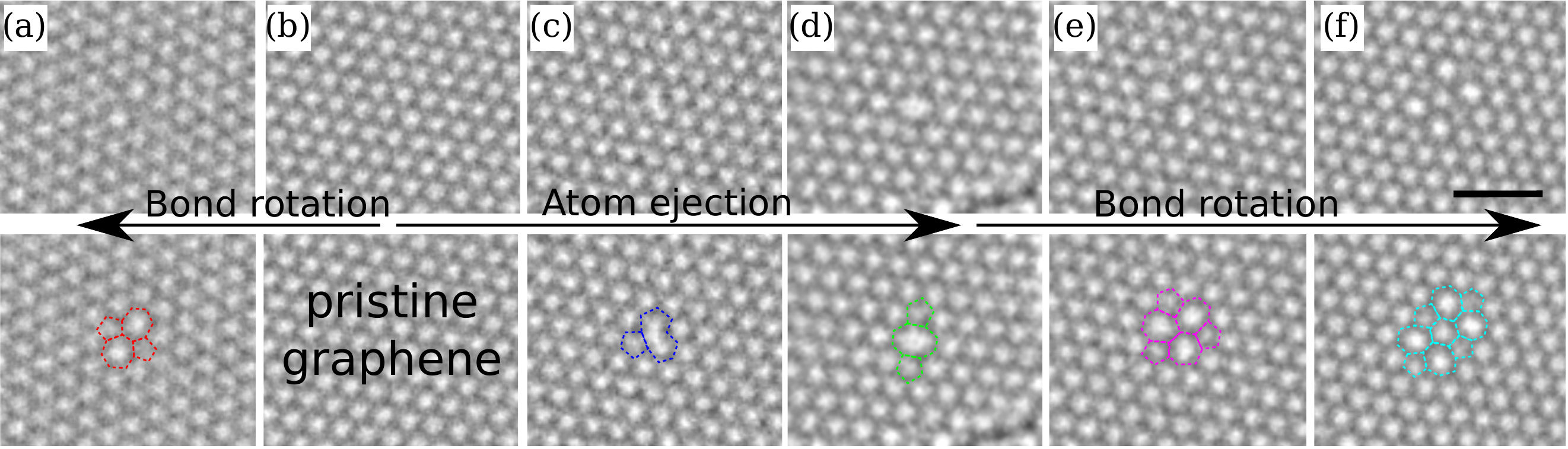} 
\caption{(Color online) Elementary defects and frequently observed
  defect transformations under irradiation. Atomic bonds are
  superimposed on the defected areas in the bottom row. Creation of
  the defects can be explained by atom ejection and re-organization of
  bonds via bond rotation. (a) Stone-Wales defect, (b) defect-free
  graphene, (c) $V_{1}$(5-9) single vacancy, (d) $V_{2}$(5-8-5)
  di-vacancy, (e) $V_{2}$(555-777) di-vacancy, (f)
  $V_{2}$(5555-6-7777) di-vacancy. Scale bar is 1nm.}
\label{fgr:Elements} 
\end{figure*}

In order to understand the mechanisms behind the transformation, we
separated them by varying the electron beam energy. To observe how a
defected graphene sheet reacts to an electron beam when atomic
ejections are prohibited by a low enough electron energy, we created
initial damage in a graphene sheet by brief 300~keV irradiation, and
then studied the generated structures at 80~keV. Now only
under-coordinated atoms can be ejected ($T_d$ for a $sp^2$--bonded C
is about 18--20~eV~\cite{Banhart1999,Smith2001}, whereas DFT
calculations predict a $T_d$ of $\sim$14~eV for a two-coordinated
C). However, bond re-organization is possible, as activation energies
for bond rotations in $sp^{2}$-bonded carbon structures are in the
range of 4--10~eV~\cite{Ewels2002,Li2005}, depending on the local
atomic configuration. Correspondingly, in pristine graphene, bond
rotations are occasionally observed under 80~keV
irradiation~\cite{Meyer2008} (Fig.~\ref{fgr:Elements}a,b), resulting
in the formation of the Stone-Wales defect~\cite{Stone1986}. In all
observed cases continued exposure reversed this transition in pristine
graphene.  However, defect structures, {\it e.g.}, di-vacancies, can
convert between different configurations (Fig.~\ref{fgr:Elements}d-f)
via bond rotations. According to our DFT simulations, the barriers for
bond rotations in these structures are 5--6~eV, which excludes
thermally activated migration.

\begin{figure*}[ht!]
 \includegraphics[width=0.72\linewidth]{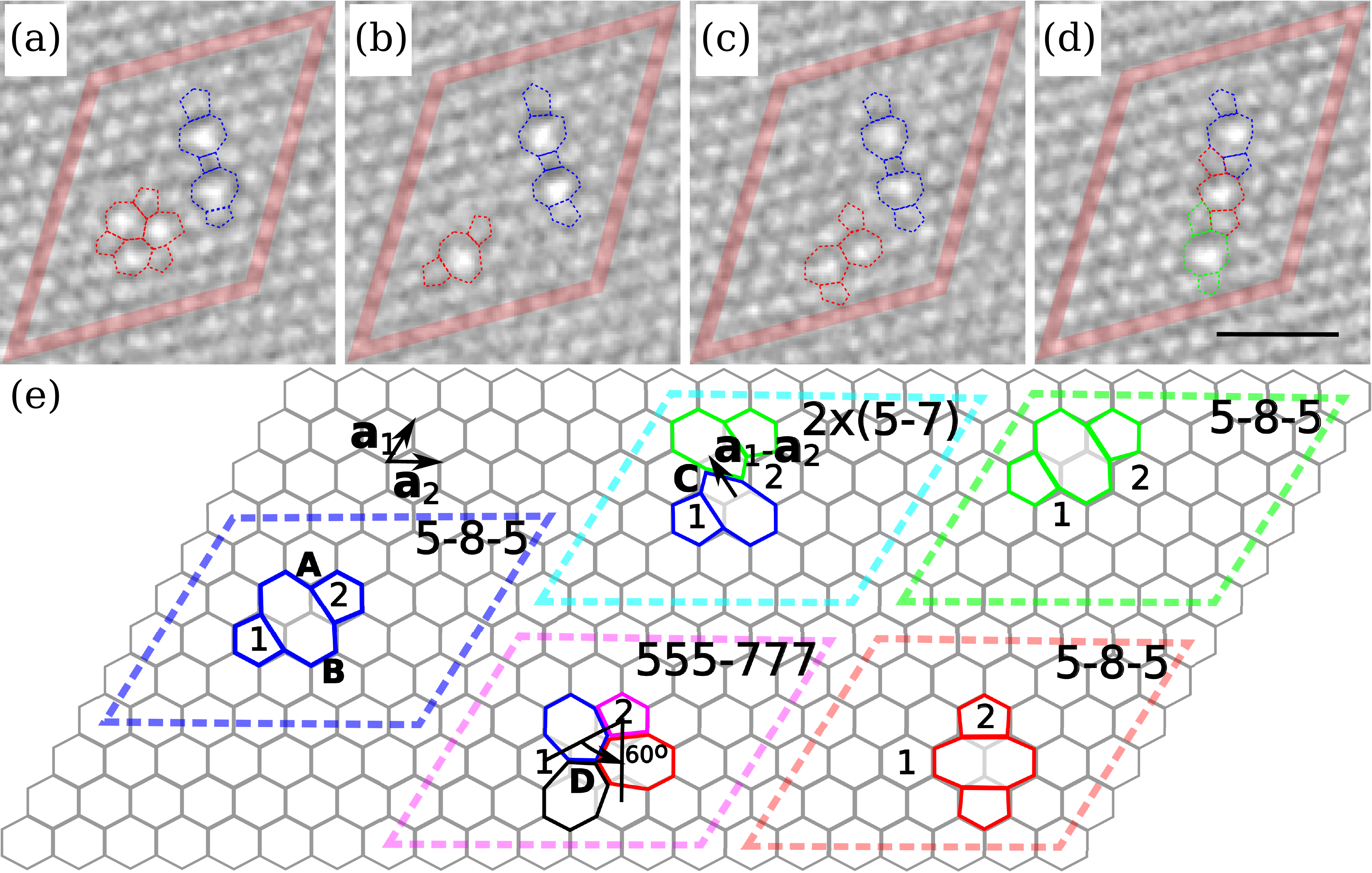}
 \caption{(Color online) (a--d) Electron beam driven di-vacancy
   migration observed at 80~keV. In (e), the changes in the bond
   configuration required for allowing the migration of a di-vacancy
   are shown.  Transformation
   $V_{2}$(5-8-5)$\rightarrow$2$\times$(5-7) is initiated by rotating
   bond A, and $V_{2}$(5-8-5)$\rightarrow$$V_{2}$(555-777) by rotating
   bond B. Rotating bonds C and D will lead to the final
   $V_{2}$(5-8-5) structures. In the first case the defect has moved
   by $\mathbf{a}_1-\mathbf{a}_2$, and in the second case it has
   rotated by 60$^\circ$ around pentagon 2. The original TEM images
   without overlays for panels (a-d) are presented in
   Ref.~\cite{Suppl}. \label{fgr:dv} }
\end{figure*}

In Fig.~\ref{fgr:dv}a--d we present evolution of a more complex defect
structure. The defects created by an electron beam are predominantly
mono-vacancies, which quickly convert to di-vacancies due to a higher
probability for under-coordinated atoms to be ejected, as noted
above. Here (Fig.~\ref{fgr:dv}a), a brief exposure to a 300~keV beam
(dose $\sim 10^{7}$~${e^{-}}/{\textrm{nm}^{2}}$) has created an
isolated $V_{2}$(555-777) and a defect with 4 missing carbon atoms
(two connected di-vacancies). During the image sequence, the 80~keV
electron beam causes the structure to re-organize via bond
rotations. The $V_{2}$(555-777) turns first into a $V_{2}$(5-8-5)
(Fig.~\ref{fgr:dv}b and then a dislocation dipole
(Fig.~\ref{fgr:dv}c), before forming a defect composed of clustered
di-vacancies (Fig.~\ref{fgr:dv}d). Fig.~\ref{fgr:dv}a--d also present
two frequently observed linear arrangements of di-vacancies.

Because atom ejection occurs at random positions, vacancies initially
appear randomly in the area exposed to the electron beam. However,
during lower-energy exposure ({\it e.g.}, 80~keV), these defects
travel via a re-bonding mechanism, which is illustrated in
Fig.~\ref{fgr:dv}e. Each migration step is initiated by a single
electron impact from which the atom obtains energy slightly below
$T_d$. In other words, electron irradiation provides the activation
energy to drive the system from a local energy minimum into another
one, in our case predominantly via (reversible) bond rotations
(Fig.~\ref{fgr:Elements}d-f, Fig.~\ref{fgr:dv}a-d). This can be
clearly seen in video S4 in Ref.~\onlinecite{Suppl} (partially shown
in Fig.~\ref{fgr:dv}a--d), where the configuration changes frequently
until it arrives in the more stable configuration composed of three
aligned double-vacancies.

Individual transitions can also lead to higher energy structures. For
example, the intermediate states for the di-vacancy migration
{[}dislocation dipole (Fig.~\ref{fgr:dv}c) and $V_{2}$(555-777)
(Fig.~\ref{fgr:dv}a)] have formation energies $E_f$ which differ from
that of the $V_{2}$(5-8-5) by $\sim +3.72$~eV and $\sim -0.66$~eV,
respectively. Aligning di-vacancies along the zigzag direction of the
lattice (Fig.~\ref{fgr:dv}d), $\sim 1.32$~eV of energy is gained per a
di-vacancy pair, as compared to isolated di-vacancies. When the
alignment appears along the armchair direction
(Fig.~\ref{fgr:dv}a--c), the energy gain is 2.01~eV. In this case, a
tetragon is formed where the two pentagons of the adjacent
di-vacancies would overlap. HRTEM simulation of the DFT-optimized
structure of the defect is in excellent agreement with the
experimental image~\cite{Suppl}. Note that $sp^{2}$-bonded carbon
tetragons in molecules, as in cyclobutadiene, can be stabilized only
at low temperatures and when the molecules are embedded into a
matrix~\cite{Legrand2010}. In our case they are stabilized by the
surrounding graphene lattice, as theoretically predicted for
nanotubes~\cite{Orellana2009}.

\begin{figure*}[ht!]
 \includegraphics[width=0.64\linewidth]{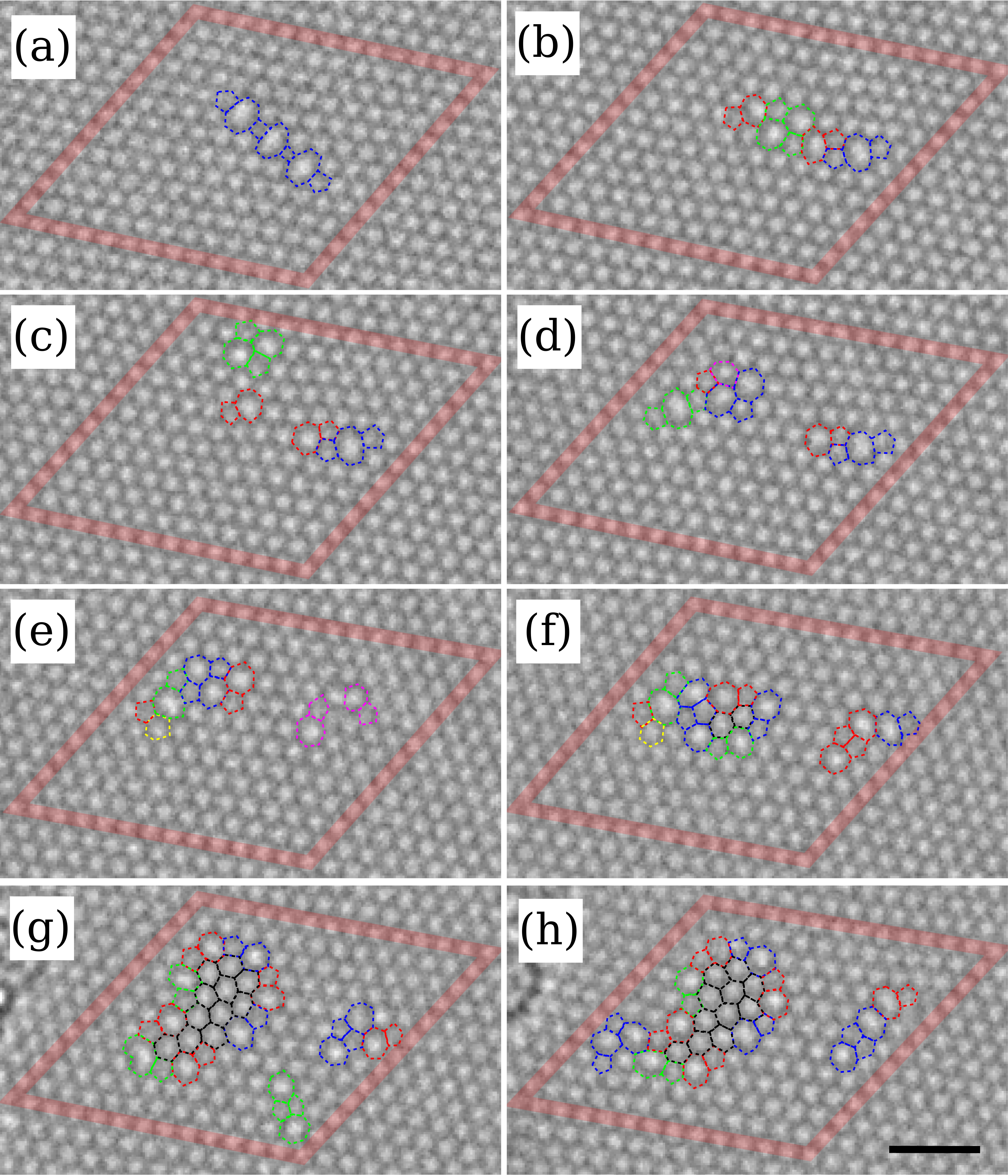} 
 \caption{(Color online) Formation of rotated hexagon-kernels in
   multi-vacancy structures under a 100~keV electron beam. Scale bar
   is 1nm. The original TEM images without overlays are presented in
   Ref.~\cite{Suppl}.\label{fgr:mvac} }
\end{figure*}

Under 100~keV irradiation, atom ejection occurs at a very slow rate,
so that changes in the atomic network are sufficiently slow to be
precisely resolved. Therefore, the reconstruction of vacancy defects
via bond rotations can be monitored immediately after a vacancy is
generated. An example is presented in Fig.~\ref{fgr:mvac}.  The
initial configuration (Fig.~\ref{fgr:mvac}a) consists of three
di-vacancies in the armchair orientation (formed prior to recording
the first image). In the recorded images, the structure loses atoms
until 24 atoms are missing. Several remarkable configurational changes
are found in this image sequence. Fig.~\ref{fgr:mvac}a--c shows a
collapse of linearly clustered defects into an apparently less
defective structure with a dislocation dipole. This corresponds to the
prediction of Jeong {\em et al.}~\cite{Jeong2008} that the dislocation
dipole is favored over a large multi-vacancy. However, this requires a
linear arrangement of vacancies. Fig.~\ref{fgr:mvac}c--d shows the
loss of four additional atoms. Two of them gave rise to an additional
di-vacancy; the other two contributed to the separation of the
dislocation cores (the rotated hexagon, clustered with a Stone-Wales
defect, constitutes a dislocation core). During continued irradiation,
we see the formation of a cluster of rotated hexagons surrounded by a
chain of alternating pentagons and heptagons
(Fig.~\ref{fgr:mvac}e--h). Such configurations, rotated by
$30^{\circ}$ with respect to the original lattice and matched by
pentagons and heptagons to the zigzag lattice direction, appear to be
the preferred way to incorporate missing atoms in the graphene
structure. Due to the matching numbers of pentagons and heptagons --
and hence cancellation of negative and positive curvature -- these
structures remain flat.

\begin{figure*}[ht!]
 \includegraphics[width=0.64\linewidth]{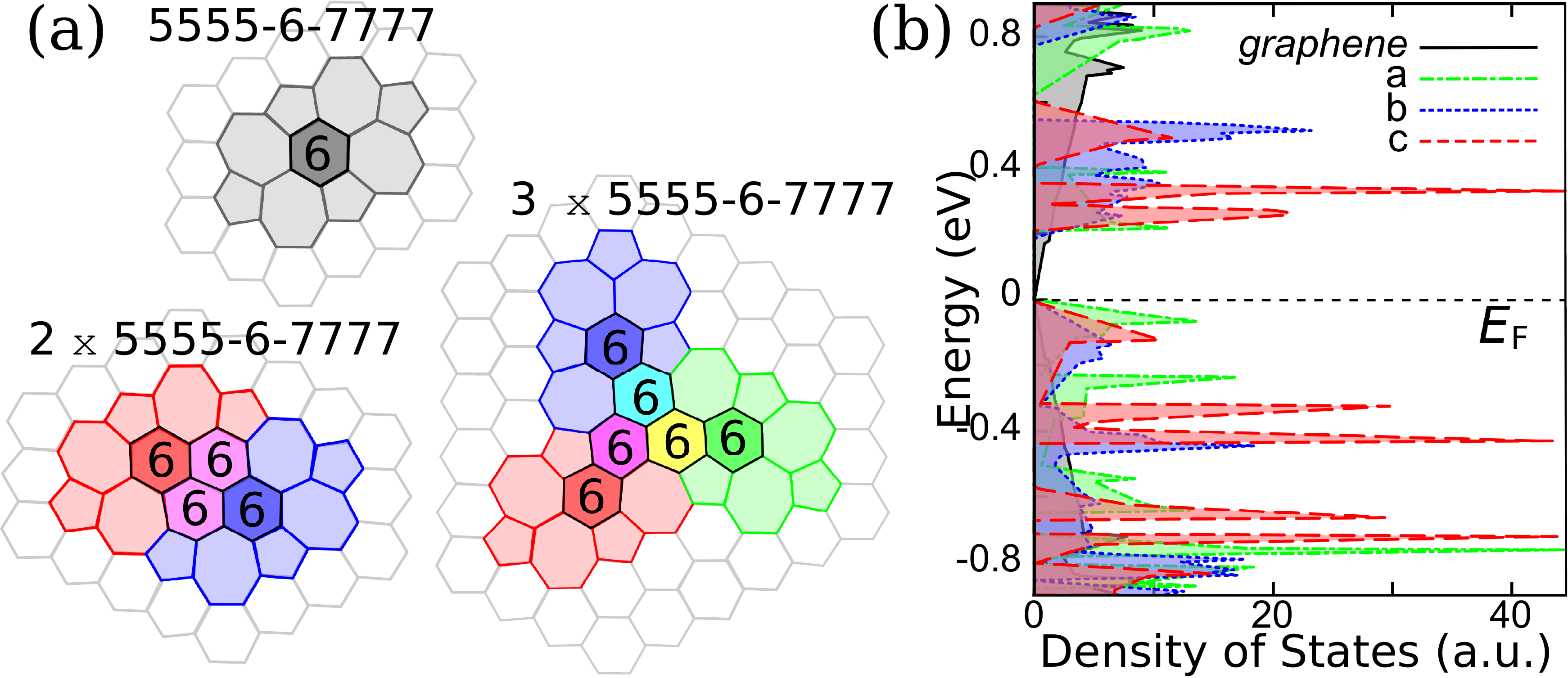} 
 \caption{(Color online) Idealized rotated-hexagon defects (a) formed
   from one, two and three $V_{2}$(5555-6-7777) di-vacancy defects, as
   optimized with DFT calculations. Electronic density of states (b)
   for pristine graphene and the structures presented in panel (a).
\label{fgr:dft}}
\end{figure*}

To understand the driving force for the transformations, we calculated
$E_f$ for the simplest defect structures matching the obverved trend
of forming a rotated hexagon kernel. The lowest-energy tetra-vacancy
(four missing atoms) can be created by combining two
$V_{2}$(5555-6-7777) di-vacancies, whereas the hexa-vacancy (six
missing atom) requires three of these defects
(Fig.~\ref{fgr:dft}a). Remarkably, these configurations have the
lowest $E_f$ of any reported vacancy structures with equal number of
missing atoms in graphene ($E_f$ per missing atom multiplied by the
number of missing atoms are $\sim 4\times$3.14~eV and $\sim
6\times$2.50~eV). A hole with six missing atoms has a formation energy
of $\sim 6\times$3.15~eV, while for a dislocation and a
h{\ae}ckelite-like structure values of $\sim 6\times$3.72~eV and $\sim
6\times$2.64~eV have been reported~\cite{Jeong2008},
respectively. Evidently, the rotated hexagon defects spawn the family
of lowest energy multi-vacancies in graphene. In contrast to what was
recently shown for the zigzag-oriented di-vacancies~\cite{Lahiri2010},
these structures open a band gap in graphene, as can be seen from the
density of states (Fig.~\ref{fgr:dft}b). The calculated band gap is in
the order of 200~meV. This value is possibly underestimated within the
used GGA approximation, and advanced DFT methods are likely to give a
higher value~\cite{Appelhans2010}.

To conclude, we have shown how an electron beam can be used to
selectively suppress and enhance bond rotations and atom removal in
graphene. We demonstrated that irradiation at electron energies just
above the threshold for atom displacement turns graphene not into a
``perforated graphene'' but a two-dimensional coherent amorphous
membrane composed of $sp^2$--hybridized carbon atoms. This membrane
grows through nucleation and expansion of defects which result from
electron beam--driven di-vacancy migration and agglomeration. These
defect configurations predominantly consist of a 30$^\circ$ rotated
kernel of hexagons surrounded by a chain of alternating pentagons and
heptagons. These defects are the energetically favored way for the
graphene lattice to accommodate missing atoms, and have a
semi-conducting nature. Since several of the presented examples of the
two-dimensional $sp^{2}$--hybridized defect configurations violate the
IPR, due to increased reactivity~\cite{Schmalz1988,Tan2009}, they may
be exploited for functionalization of graphene. We also showed
unambiguous evidence for four-membered carbon rings in graphitic
structures. Clearly, despite of the large amount of research, the
richness of carbon chemistry continues to provide surprises. More
examples of the observed structures and videos of the complete TEM
image series are presented in Ref.~\cite{Suppl}.

We acknowledge support by the German Research Foundation (DFG) and the
German Ministry of Science, Research and the Arts (MWK) of the state
Baden-Wuerttemberg within the SALVE (sub angstrom low voltage electron
microscopy) project and by the Academy of Finland through several
projects. We are grateful for the generous grants of computer time
provided by CSC Finland.

\section*{Methods}

As described in the main article, graphene membranes were prepared by
micro-mechanical cleavage and transfer to TEM
grids~\cite{Meyer2008a}. Aberration-corrected HRTEM imaging was
carried out in an FEI Titan 80--300, equipped with an objective-side
image corrector. The microscope was operated at 80~keV and 100~keV for
HRTEM imaging, and at 300~keV for irradiation.  The extraction voltage
of the field emission source was set to a reduced value of 2~kV in
order to reduce the energy spread. For both 80~keV and 100~keV
imaging, the spherical aberration was set to 20~$\mu$m and
images were obtained at Scherzer defocus (ca. $-9$~nm). At these
conditions, dark contrast can be directly interpreted in terms of the
atomic structure. Image sequences were recorded on the CCD camera with
exposure times ranging from 1~s to 3~s and intervals between 4~s and
8~s, and a pixel size of $0.2~\textrm{\AA}$. The effect of slightly
uneven illumination is removed by normalization (division) of the
image to a strongly blurred copy of the same image, effectively
removing long-range variations. Drift-compensation is done using the
Stackreg plugin for the ImageJ software \cite{Thevenaz1998}. We show
individual exposures as well as averages of a few frames (up to 10).
This is because we have used different beam current densities, in
order to test for possible dose rate effects (within the 100~keV,
simultaneous imaging and defect generation experiment). The
configurations described here can be discerned in individual exposures
if ca. $10^{4}$ counts per pixel at $0.2~\textrm{\AA}$ pixel size are
used. With a high beam current density, this is possible in 1~s
exposures (corresponding to a total dose of
ca. $10^{5}\frac{e^{\textrm{-}}}{\textrm{\AA}^{2}}$ per image, and a
dose rate of $10^{5}\frac{e^{\textrm{-}}}{\textrm{\AA}^{2}\cdot s}$).
At lower current densities, the same dose was spread over several
exposures, so that sample drift could be compensated (individual
exposures could not be longer than 3~s due to sample drift). The
lowest dose rate was
ca. $3\cdot10^{3}\frac{e^{\textrm{-}}}{\textrm{\AA}^{2}\cdot s}$.
Within our dose rate range, density and shape of defects appears to
depend only on the total dose.

Formation energy of a structure in our DFT calculations was defined in
the usual way as
\begin{equation}
  E_{f}(V_{n})=E_{\mathrm{tot}}(V_{n})-E_{\mathrm{tot}}^{\mathrm{gr}}+n\mu_{\mathrm{gr}},\end{equation}
where $E_{\mathrm{tot}}(V_{n})$ and $E_{\mathrm{tot}}^{\mathrm{gr}}$
are the total energies of the structure with the defect ($n$ missing
atoms) and the same supercell without the defect, respectively, and
$\mu_{\mathrm{gr}}$ is the chemical potential of a carbon atom in
pristine graphene. The semi-conducting features of the defects were
consistently observed with varying number of $k$-kpoints used in
calculating the electronic density of states.

  \section*{Further images}

\begin{figure*}
\includegraphics[width=0.8\linewidth]{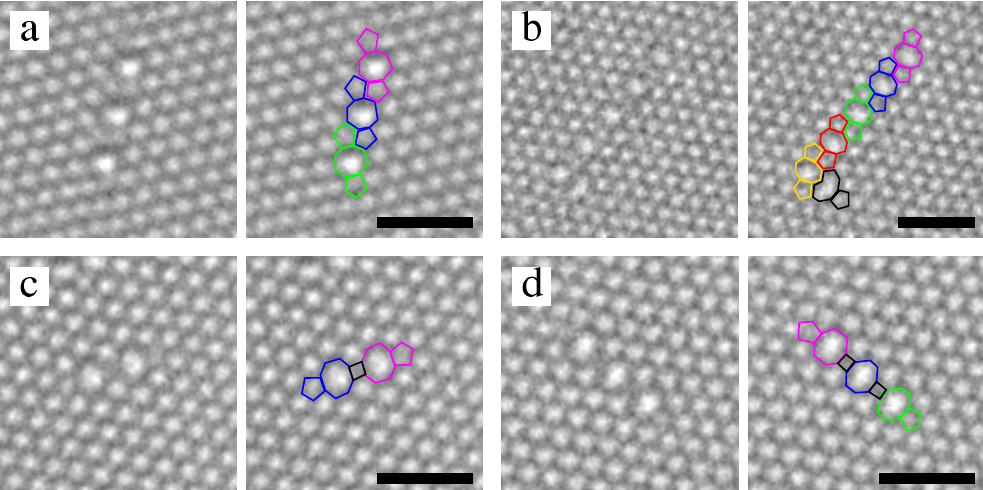}\caption{Images of di-vacancy (DV)
  defects in linear alignment, shown with and without structure
  overlay in each case. (a) Three DVs aligned along the zigzag
  direction of the graphene lattice. (b) Five DVs aligned in zigzag
  direction, clustered with a single vacancy (black). (c) Two DVs
  aligned along the armchair direction, forming a carbon tetragon at
  their intersection. (d) Three DVs aligned along the armchair
  direction, forming two carbon tetragons. All scale bars are 1~nm. }
\label{px::dv} 
\end{figure*}

\begin{figure*}
\includegraphics[width=0.6\linewidth]{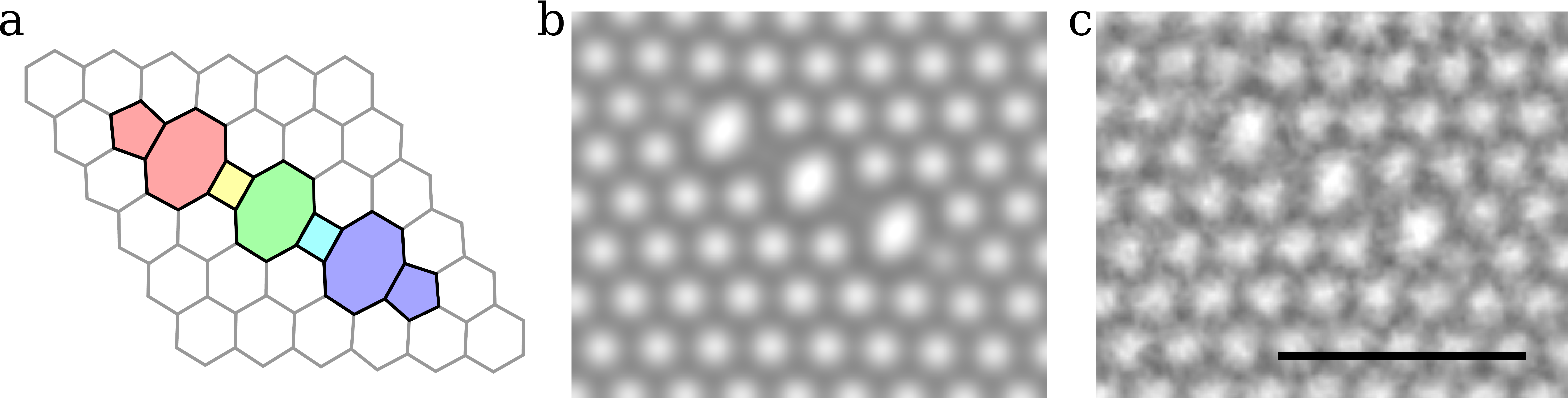}\caption{DFT-optimized
  structure of the defect with two tetragons (a), HRTEM simulation
  based on the DFT-structure (b) and the HRTEM image of the same
  structure for comparison (c). Scale bar is 1~nm.}
\label{px::dv2} 
\end{figure*}

  For clarity we show here the TEM images of the main article without
  stucture overlay, and also a few additional TEM
  images. Fig.~\ref{px::dv} shows several di-vacancy (DV) defects in
  linear alignment. In particular, the carbon tetragon is always
  reproduced when the DVs are aligned along the armchair direction of
  the graphene lattice. Fig.~\ref{px::dv2} shows a comparison between
  a HRTEM image simulation based on a DFT-optimized structure and an
  actual HRTEM image for the armchair alignment of di-vacancies, which
  contains the carbon tetragons.

\begin{figure*}
\includegraphics[width=0.6\linewidth]{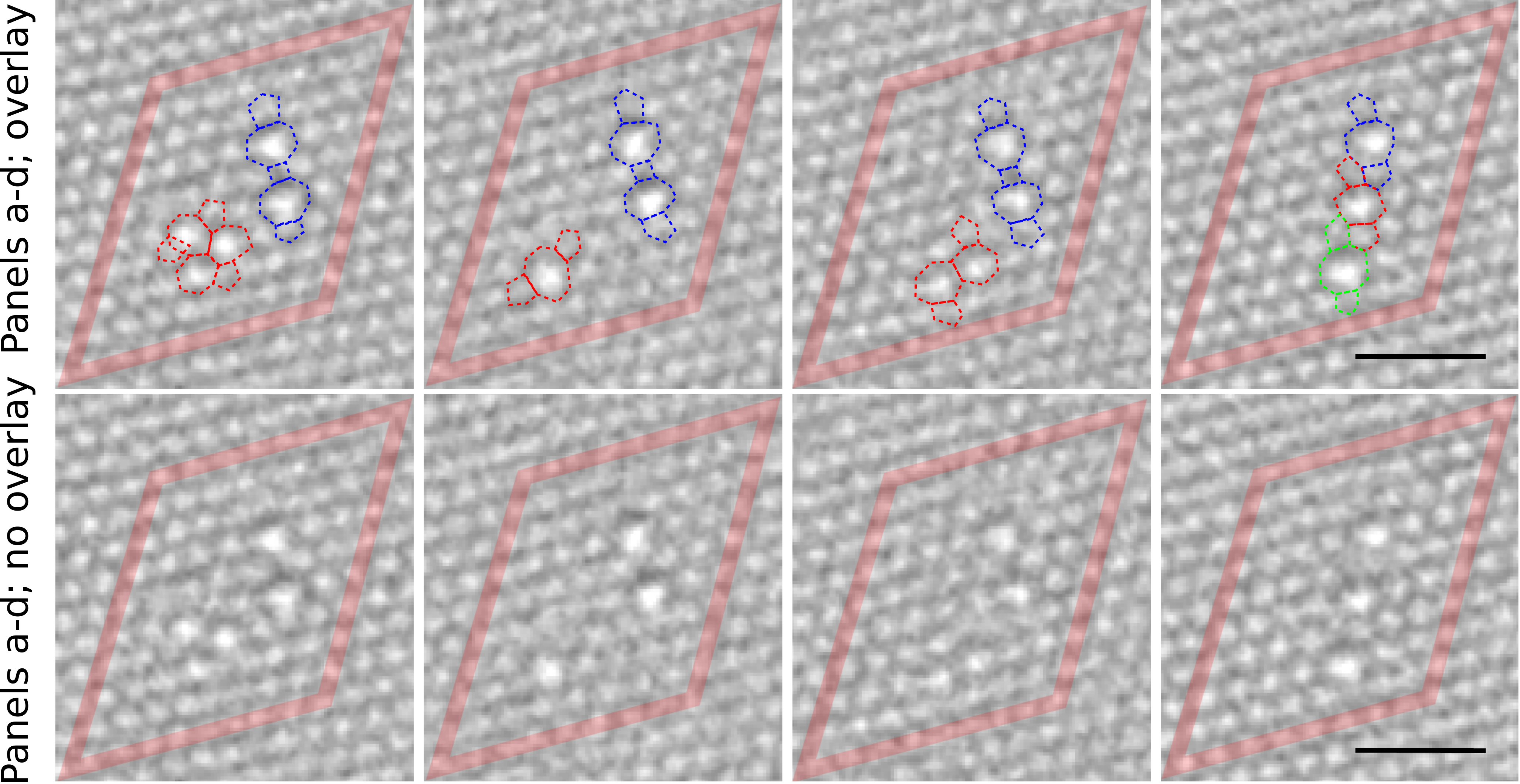}\caption{Panels a--d from
  Figure~3 of the main article with and without overlays.  Scale bar is
  1~nm.}
\label{px::sfig2} 
\end{figure*}

\begin{figure*}
\includegraphics[width=1\linewidth]{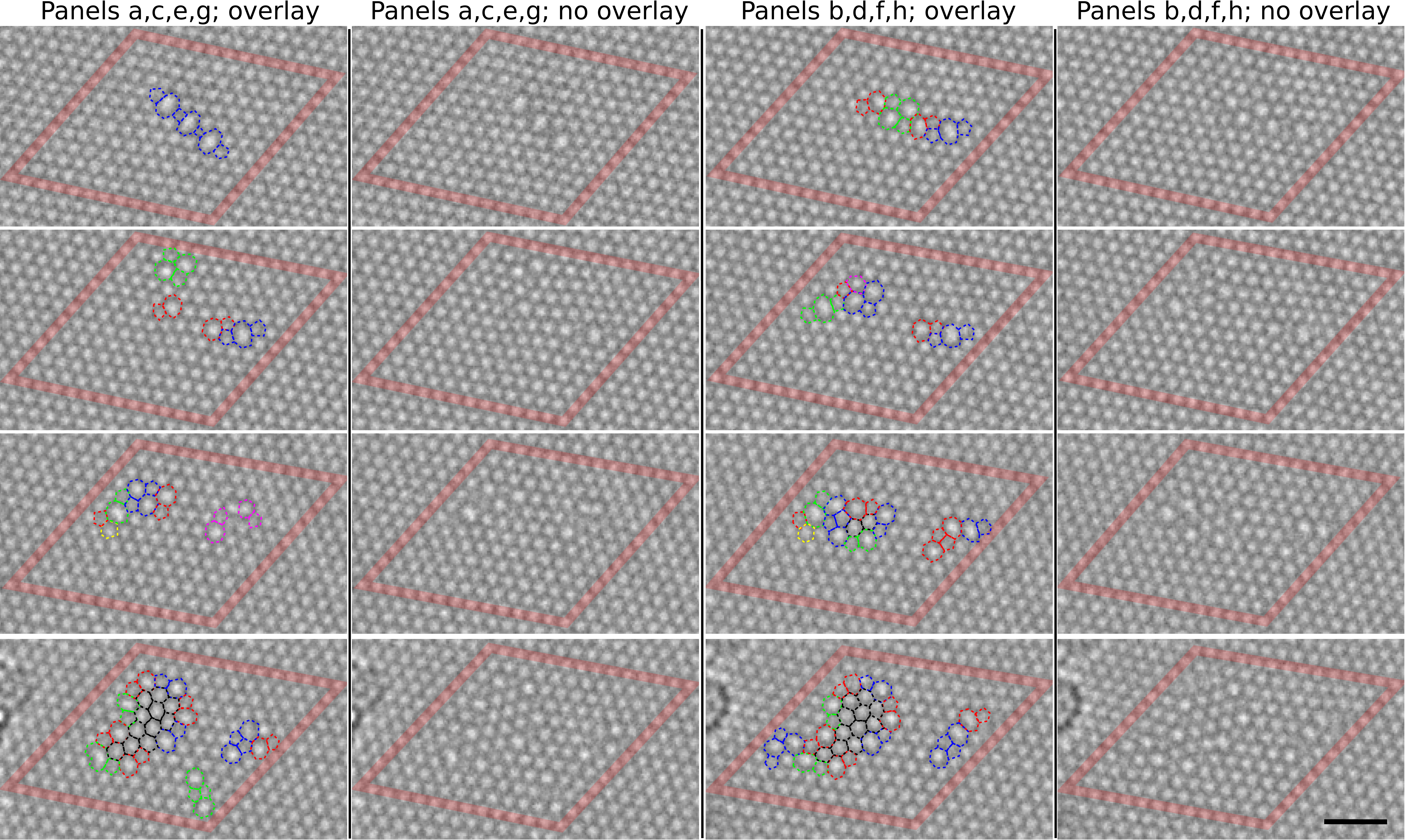}\caption{Panels a--h from
  Figure~4 of the main article with and without overlays.  Scale bar is
  1~nm.}
\label{px::sfig3} 
\end{figure*}

In Fig.~\ref{px::sfig2}, we show the panels a--d from the Figure~3 of
the main article with and without overlays. These images are also
contained in Supplementary video S2. In Fig.~\ref{px::sfig3} we show
panels a--h from the Figure~4 of the main article with and without
overlay (the corresponding time series is shown in Supplementary video
S3).

In Fig.~\ref{px::300kV}, we show two additional images where a defect
with a rotated hexagon kernel was generated from clusters of multiple
vacancies.  These configurations were created by a short 300~keV
exposure and subsequent imaging at 80~keV. It should be noted that the
rotated hexagon kernels appear in the larger vacancy clusters in both
of our experiments (100~keV irradiation with simultaneous imaging as
in Figure~4 of the main article, and the 300~keV/80~keV combination as
shown in Fig.~\ref{px::300kV}).

\begin{figure*}
\includegraphics[width=0.5\linewidth]{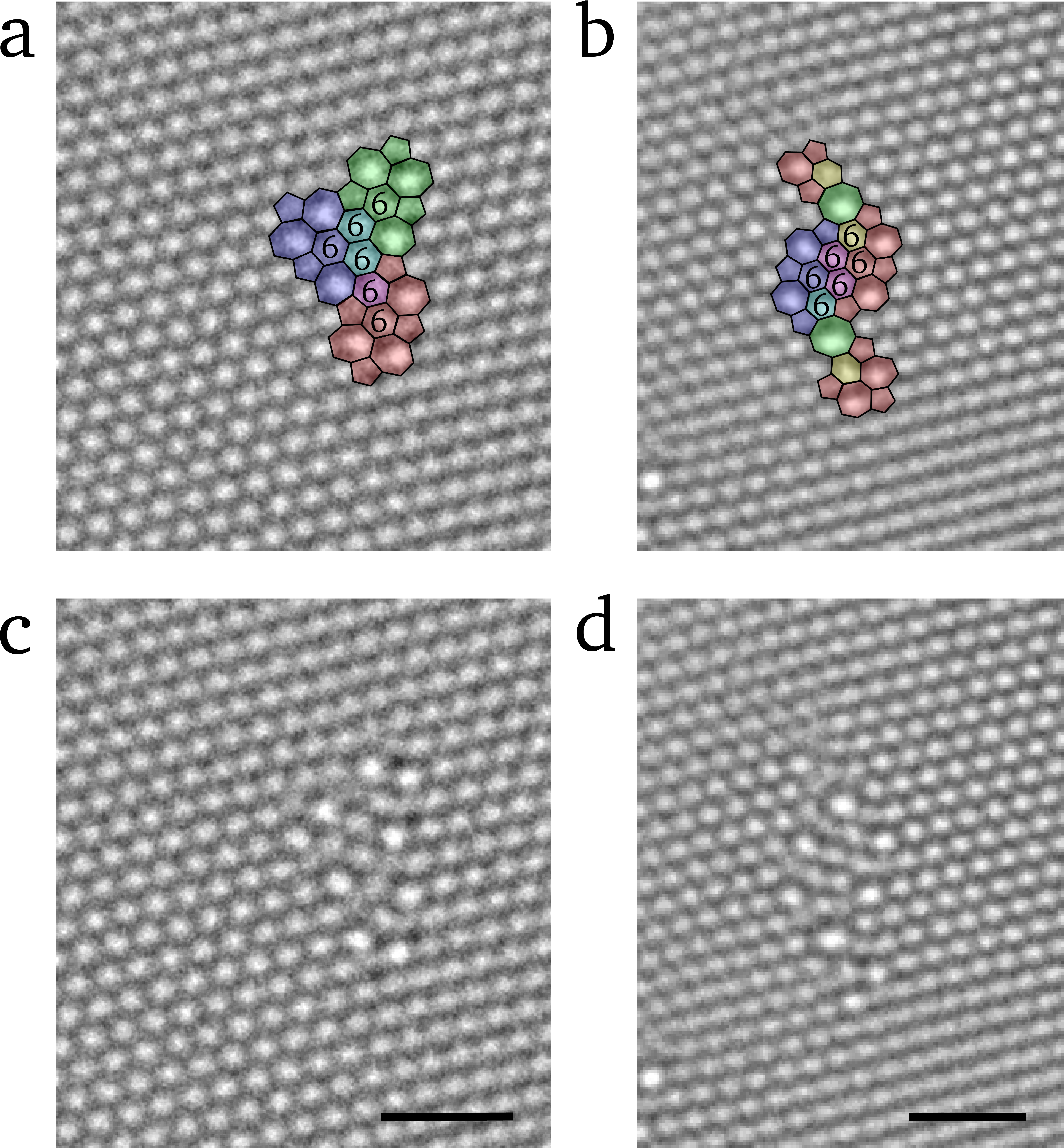} \caption{Two
  examples of defect structures created by short 300~keV exposure and
  imaged at 80~keV. In the upper row (panels a,b) DFT-optimized
  configurations are superimposed over the experimental images. Lower
  row (panels c,d) displays the structures without DFT-overlays. The
  first structure (a,c) can be constructed by combining three
  $V_{2}$(5555-6-7777) reconstructed di-vacancies (6 missing atoms),
  whereas the second one (b,d) can be formed from two
  $V_{2}$(5555-6-7777)'s, two $V_{2}$(5-8-5)'s, one $V_{2}$(555-777)
  and one (55-77) defect (total of 10 missing atoms), as indicated by
  the colors. Scale bars are 1~nm.}
\label{px::300kV} 
\end{figure*}

\section*{Description of the supplementary videos}

\noindent {\em Supplementary video S1}: Generation and transformation
of defects under 100~keV irradiation. Each frame is an average of 2
CCD exposures. Note the continuous increase in defect density and
partial amorphization of the membrane. \\

\noindent {\em Supplementary video S2}: Further transformation of
graphene to a 2D amorphous membrane under 100~keV irradiation.  In
terms of total dose, it can be considered as a continuation of video
S4 (although sample region is different). Each frame shows an
individual CCD exposure. Note that the entire clean graphene area
becomes amorphous while remaining one-atom thick, and beam-generated
holes make up only a small fraction of the area. \\

\noindent {\em Supplementary video S3}: Two isolated di-vacancies,
generated by brief 300~keV irradiation and observed at 80~keV. The
di-vacancies migrate under the beam and transform between the
$V_{2}$(5-8-5), $V_{2}$(555-777) and $V_{2}$(5555-6-7777)
configurations multiple times. The video shows individual exposures in
each frame.\\

\noindent {\em Supplementary video S4}: A cluster of several vacancy
defects, generated by brief 300~keV irradiation and observed at
80~keV.  The video shows individual exposures in each frame. The
configuration changes continuously until the final, linear aligned
di-vancy configuration is observed and stable throughout several
exposures (frames 22--29).  Frames 30--36 are duplicates of frame 29
in order to show the final configuration as a still image at the end. \\

\noindent {\em Supplementary video S5}: Generation and transformation
of defects under 100~keV irradiation. Each frame is an average of 10
CCD exposures (recorded at a lower current density). This video shows
the generation of the rotated hexagon kernel as in Figure~4 of the main
article. The rotated hexagon kernel is highlighted as overlay in
frames 28--31.

\end{document}